%% file: ieee_software_2010.tex
\documentclass[epj]{svjour}
\usepackage[pdftex]{graphicx}
\usepackage{listings}
\usepackage{mdwlist}
\graphicspath{{figures/}}
\DeclareGraphicsExtensions{.eps,.pdf,.jpeg,.png}

\begin{document}

\title{The Architecture of MEG Simulation and Analysis Software}

\author{Paolo~W.~Cattaneo\inst{1} \and 
Ryu~Sawada\inst{2}\and
Fabrizio~Cei\inst{3}\and
Shuei~Yamada\inst{4}\and
Matthias~Schneebeli\inst{5} 
\thanks{\emph{Present address:} DECTRIS Ltd., Neuenhoferstrasse 107, CH-5400 Baden, Switzerland} 
}

\institute{INFN Pavia, Via Bassi 6, Pavia, I-27100, Italy 
          \and ICEPP, The University of Tokyo, 7-3-1 Hongo, Bunkyo-ku, Tokyo 113-0033, Japan 
          \and INFN and Department of Physics of the University of Pisa, Largo B.Pontecorvo 3, Pavia, I-56127, Italy 
          \and KEK, High Energy Accelerator Research Organization, 1-1 Oho, Tsukuba, Ibaraki 305-0801, Japan
          \and Paul Scherrer Institute PSI, CH-5232 Villigen, Switzerland and
        Swiss Federal Institute of Technology ETH, CH-8093 Z\"urich, Switzerland}








\date{04 March 2011}
\input{abstract}

\maketitle



\section{Introduction}
\input{introduction}

\section{The MEG software structure}
\input{structure}

\section{REM: a FORTRAN 77 framework }
\input{rem}

\section{GEM: the Monte Carlo simulation }
\input{gem}

\section{The database}
\input{database}

\section{ROME: a framework generator }
\input{rome}

\section{Readout simulation and event mixing}
\input{bartender}

\section{The reconstruction and analysis program}
\input{analyzer}

\section{Offline processing}\label{sec_offline}
\input{offline}

\section{Conclusion}
\input{conclusion}


\section*{Acknowledgment}
We acknowledge the role of Dr.~Stefan~Ritt from PSI,
who is the main author of the online software MIDAS.\\
Integration of each sub-detector part was done by many collaborators; forty
of them have contributed to the MEG software.



\bibliographystyle{unsrt}
\bibliography{meg}





\end{document}

%% file: abstract.tex
\abstract{
\PACS{
     {29.85.Fj}{Data analysis}
    } 
MEG ($\mu^+ \rightarrow e^+\gamma$) is an experiment dedicated to search for the 
$\mu^+ \rightarrow e^+\gamma$ decay that is strongly suppressed in the Standard Model, but
allowed in many alternative models and therefore very sensitive to new physics. 
The offline software is based on two frameworks. The first is {\bf REM} in FORTRAN 77, which
is used for the event generation and detector simulation package {\bf GEM}.
The other is {\bf ROME} in C++, used for the readout electronics simulation 
{\bf Bartender} and for the reconstruction and analysis program {\bf Analyzer}.
Event display in the simulation is based on GEANT3 graphic libraries and in
the reconstruction on ROOT graphic libraries.
Data are stored in different formats at various stages of the processing.
The frameworks include utilities for I/O, database access and format conversion 
transparent to the user.
}

%% file: introduction.tex
\begin{figure}[htbp]
  \begin{center}
    \includegraphics[width=.70\linewidth]{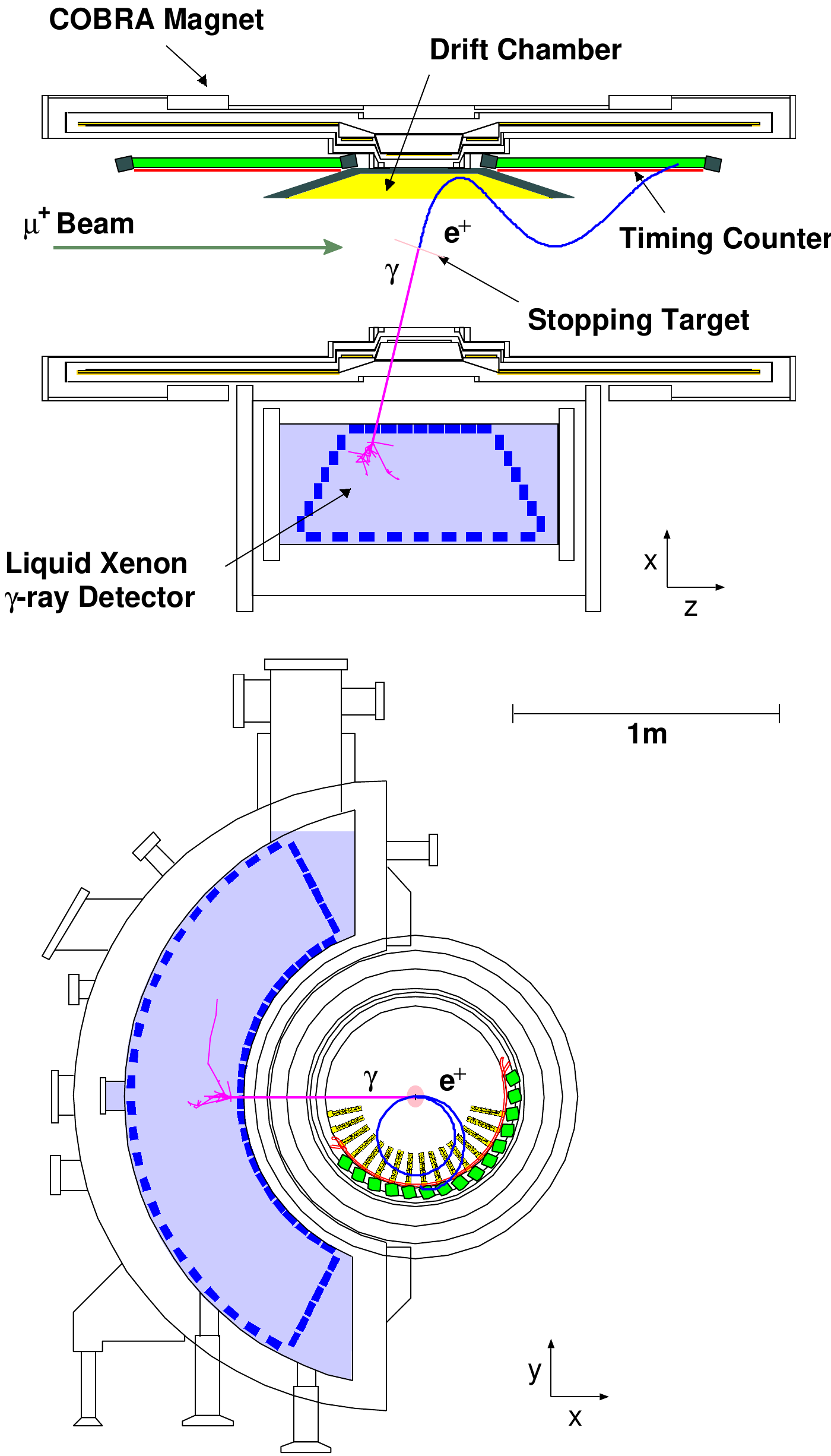}
    \caption{The MEG experimental setup.}
    \label{setup}
  \end{center}
\end{figure}

The MEG experiment at Paul Scherrer Institute (PSI) in Switzerland is searching 
for the rare decay $\mu^+\rightarrow e^+\gamma$, employing a very intense 
($3\times 10^7 \mathrm{s}^{-1}$) $\mu^+$ beam, which is stopped in a thin target at 
the center of the detector. MEG is a small-size collaboration 
($\approx 50-60$ physicists at any time) with a life span of about 10 years.\\
The collaboration started the software development in 2002 after a few years of 
prototype studies, with the goal of being ready for data taking in a technical run 
foreseen after 3 years.
Since the beginning, the tight time schedule and the limited human resource 
available, in particular in the offline architecture group, emphasized the 
importance of reusing software developed during the prototype studies and 
exploiting existing expertise.
Therefore great care has been devoted to provide a simple system that hides 
implementation details to the average programmer. That has allowed many members 
of the collaboration with limited programming skill to contribute to the 
development of the software of the experiment.\\
The detector consists of a Liquid Xenon Calorimeter for measuring the $\gamma$ 
momentum vector and timing and of a spectrometer consisting of a set of 
drift chambers and of a timing counter embedded in a strong gradient magnetic 
field generated by a superconducting magnet (COBRA) 
for the measurement of $e^+$ kinematic variables. A sketch of the apparatus is in 
Fig.\ref{setup}. The waveforms from readout electronics are digitized at 
$\approx 1\,\mathrm{GHz}$ frequency and stored in the output to optimize 
time resolution \cite{Ritt2010486}.\\ 
Waveform data is encoded in a format developed in the MEG group.
The data of each channel consists of a header and binary waveforms.
Each header contains a hardware channel number and parameters needed to decode data.
The data can be encoded in different ways depending 
on required compression factor, precision and characteristics of waveforms 
of each subdetector. 
The experiment totals $\approx 3000$ channels and reduction by a factor 
of 3 in data size is achieved applying zero suppression, waveform resampling 
or restricting the recorded region depending on the subdetector.\\ 
The typical DAQ event rate is $\approx 6$ Hz. 
Data size is about 4.8 GB per run for 2000 events. Data files are compressed in the offline-cluster
by a factor of 2. Event size after the compression is 1.3 MB/event.\\
During $\approx 3$ months in 2010, $\approx 21\times 10^6$ $\mu^+\rightarrow e^+\gamma$ 
triggers were collected for a total of 60 TB of data written on disk, 
half of which from physics runs and the rest from calibration runs.\\
The software requirements include the simulation of the generation of signal 
and background events, of their interaction with the detector and of the read out,
the reconstruction from raw data, real or simulated, to high level objects, 
e.g. tracks and photons as well as providing an analysis environment.\\
The average time for simulating the interaction of a signal event in the detector is
$5.8$ s/event, while the average time for simulating the readout electronics is
$1.2$ s/event. The average time for reconstruction is $1.6$ s/event.\\
The software organization designed to comply with these requirements is presented.

%% file: structure.tex

\begin{figure*}[htbp]
  \begin{center}
    \includegraphics[width=.95\linewidth]{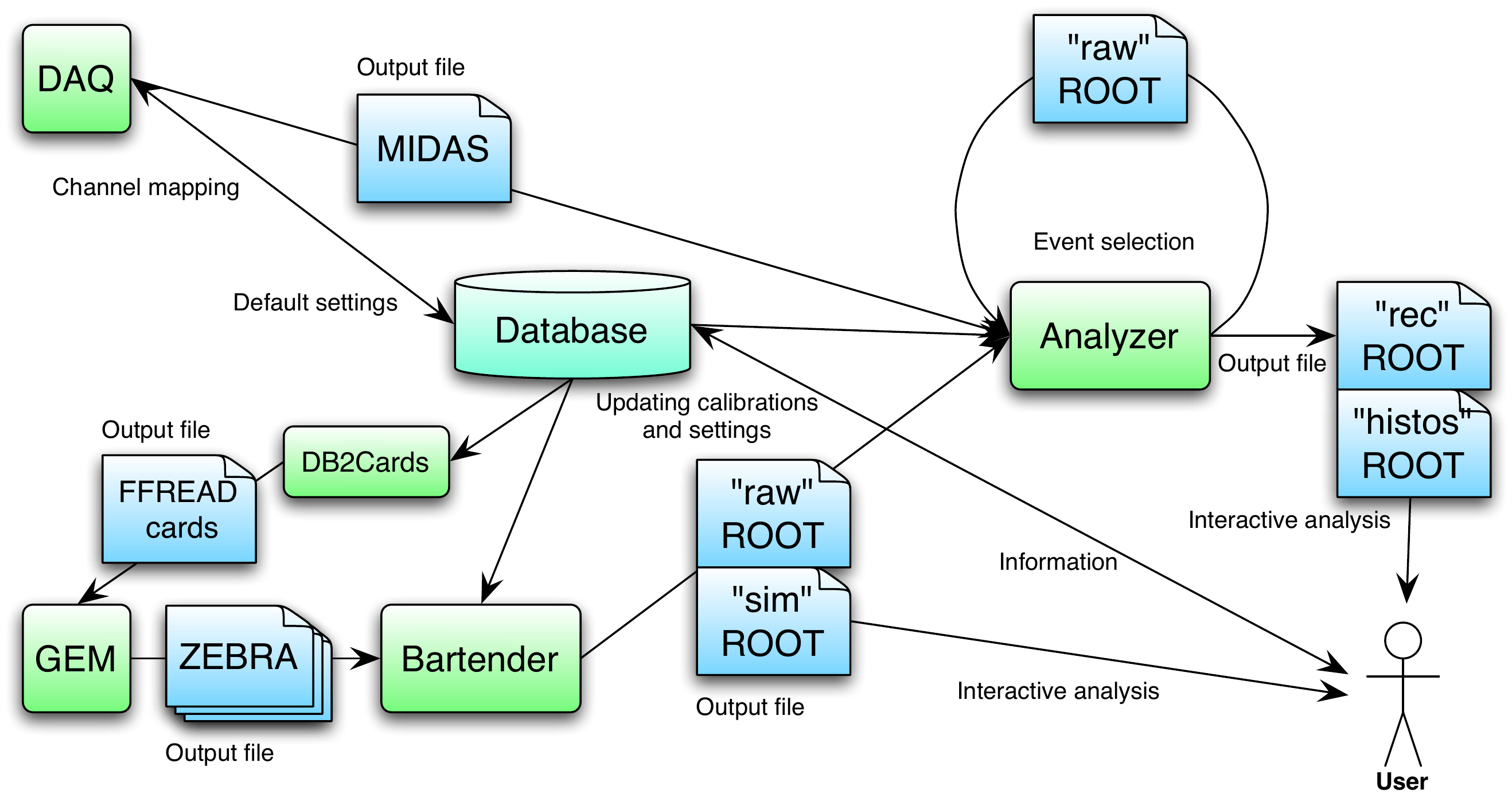}
    \caption{Connection between MEG software components}
    \label{meg_software}
  \end{center}
\end{figure*}

The MEG offline software consists mainly of GEM (event-generation, particle tracking and detector
simulation), Bartender (event-mixing and electronics simulation) and Analyzer (recostruction and analysis
of experiment and simulation data). The relations between the various software components   
are shown in Fig.~\ref{meg_software}.\\
The parameters in use in the programs are managed through a common SQL database.\\
The MEG DAQ system is based on MIDAS \cite{midas}; raw experimental data are therefore
saved as binary files in the native format of the system. 
Briefly the MIDAS format consists in an event header followed by MIDAS banks. Each bank 
is defined by a 4 character name and contains a description of the unique data type and 
an array of data.\\
The DAQ system inserts run information and default analysis parameters into the
database when a run is taken.\\
These files are read by Analyzer that reconstructs the events and produces two files:
a \verb+rec+ ROOT \cite{root},\cite{Brun199781} file, which contains a ROOT \verb+Tree+ and a 
histogram file for quick data checks.
Before Analyzer starts a run, analysis parameters are read from the database. The analysis
parameters (geometry, calibration etc.) can be changed later by users and data can be
reprocessed with the updated parameters.
If necessary, Analyzer copies raw data of selected events (cut for physics analysis) into \verb+raw+ ROOT files for
future reprocessing.\\
The simulation program GEM, steered by configuration files created by the 
DB2Cards program by reading the database, generates various types of events 
that are propagated in the detector. It is based on GEANT3 and CERNLIB and 
outputs data in exchange ZEBRA format \cite{zebra}.
Bartender reads those files and simulates the readout electronics to convert 
hits into waveforms. The simulated waveforms are written in \verb+raw+ ROOT files whose 
bank structure is the same as experimental data in MIDAS files.
In \verb+sim+ files simulation specific variables, such as kinematics of generated
particles and true hit information, are saved. Analyzer reconstructs
events from \verb+raw+ files using the same algorithms as for the experimental data.
High level physics analysis is also realized within Analyzer.\\
Version control is managed by the Subversion \cite{svn} package.

%% file: rem.tex

As anticipated above, the technical choices in designing the offline architecture 
were driven by considerations about the time schedule, the man power and the technical
skills available in the collaboration at the start of the project. The existence of 
important fragments of simulation code developed in FORTRAN 77 and GEANT3 
during the prototype phase at the time of the choice motivated the collaboration 
to retain the programming language and the library for the simulation of the experiment.\\
Nevertheless the simulation software was organized following a modern programming 
paradigm, that is using an Object Oriented approach organized in a framework 
\cite{Brunepjp}.

\subsection{Implementation of a FORTRAN 77 framework}

The detector simulation section GEM of the MEG software is written in FORTRAN 77, that
was designed for procedure oriented structured programming, not for OO programming.\\
Nevertheless a programming paradigm can be implemented in a variety of programming 
languages, even not designed for it. A limited but satisfactory support to the OO 
paradigm is at reach also in FORTRAN 77 on the basis of the following list of 
approximate equivalences between procedure oriented and OO concepts 

\begin{itemize}
\item Class $\leftrightarrow $ Library
\item Class data $\leftrightarrow $ Data structure (FORTRAN 77 Common block)
\item Class interface $\leftrightarrow $  Set of library routines 
\item Base Class $\leftrightarrow $ Module standardization
\item Virtual Class $\leftrightarrow$ Alternate choice of libraries
\end{itemize}

\subsection{Modules}

The Module is the basic unit manipulated by the framework that corresponds
to an OO class. Each Module is implemented concretely in a library.
There are different types of Modules, that can be classified as

\begin{itemize}
\item Basic Module : empty Module
\item Steerable Module : Module steerable by configuration files (cards)
\item Data Module : contains only data
\item Algorithm Module : implements an algorithm using other Modules
\item Service Module : provides interface to external libraries
\end{itemize}

These types share a common set of routines and differ by additional functionalities 
depending on the Module type implementing the OO paradigm of class hierarchy.

\subsection{The framework: REM}

The framework is a Module with an event loop.
The Modules associated to the framework are accessed in sequence by calling 
their routines in the corresponding framework routines.\\
Three module are provided by default in REM
\begin{itemize}
\item Steering cards: FFREAD package
\item I/O : ZEBRA I/O
\item Histogramming : HBOOK package
\end{itemize}

The others Modules are project dependent and their routines are called in
the corresponding framework user routines. 
These user routines, provided empty by default, are called by the framework routines. 
They can be overwritten implementing the OO inheritance mechanism.

%% file: gem.tex
\begin{figure}[htbp]
  \begin{center}
    \includegraphics[width=.95\linewidth]{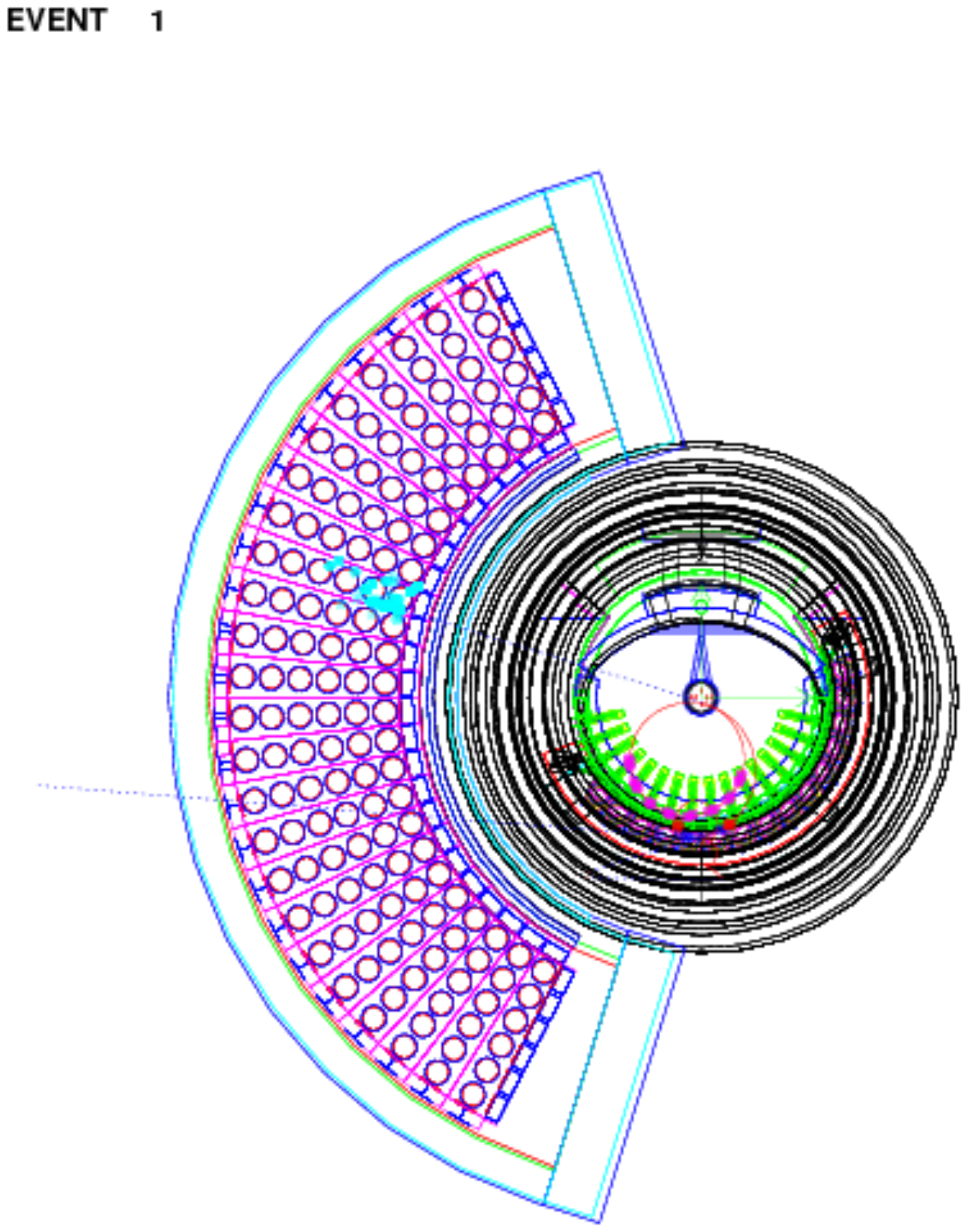}
    \caption{A $\mu^+\to e^+\gamma$ simulated event: the $e^+$ track is in red with hits in 
     drift chambers and timing counter in violet, red and blue,
     the $\gamma$ track in blue with hits in Liquid Xenon Calorimeter in cyan.}
    \label{gem_event}
  \end{center}
\end{figure}

The propagation of the $\mu^+$ beam in the last section of the beam line, its 
interaction in the target, the particle decay and the propagation and interactions
of the decay products in the detector are simulated with a FORTRAN 77 Monte Carlo 
program (GEM) based on the GEANT3 package \cite{GEANT3}. 
GEM can generate several event types, such as
$\mu^+\rightarrow e^+\gamma$ signal (shown in the Fig.~\ref{gem_event}), radiative
muon decay, Michel muon decay, cosmic ray, alpha source calibration and many others.
GEM incorporates a detailed 
description of the material and simulates the interactions of the particles in the 
detector as well as the response of the sub-detectors up to the readout stage. 
In particular the photon propagation in the Liquid Xenon Calorimeter and in the 
Timing Counter is simulated in detail.\\
The program is heavily modularized using the FORTRAN 77 framework REM. 
This approach simplifies the addition of new Modules; 
Modules can be either sub-detector simulation sections or 
service tools like e.g. graphics.\\
Within this approach, the GEANT3 library can be treated as a Module and sequenced
like any other module.\\
GEM is steered by configuration files, called cards, read by the FFREAD package 
\cite{FFREAD}, that is available in REM. These cards can be generated 
through the DB2Cards that is a ROME based framework. 
DB2Cards reads parameters from the database and 
output FFREAD cards, one for each Module, under the control of a XML 
configuration file. This file permits to select the simulation configuration, e.g.
year dependent or calibration setups, that are maintained in the database.\\ 
The most natural choice for the format of the GEM output files is ZEBRA. 
Potential disadvantages of this approach are that the manipulation of ZEBRA banks
is not user friendly, error prone and requires significant knowledge of the package.
A solution to these problems consists in manipulating only variables in common blocks
in the code and then mapping these variables into the output ZEBRA banks.
That is done automatically by providing a bank description based on the DZDOC format 
\cite{zebra} and generating through a Perl \cite{perl} script the following routines
for each bank \verb+xxxx+

\begin{basedescript}{\desclabelstyle{\pushlabel}\desclabelwidth{6em}}
\item[get\_xxxx]   Fetch the bank link
\item[print\_xxxx] Print out of the bank
\item[build\_xxxx] Fill the bank with the variables in common block (before writing out)
\item[fill\_xxxx]  Fill the common block variables with bank content (after reading in)
\end{basedescript}

\noindent GEM provides for each Module \verb+yyyy+ the routines \verb+fill(build)yyyyrunheader+ 
and \verb+fill(build)yyyyeve+ that call all corresponding routines of the banks related to 
the module. GEM provides also the routines \verb+fill(build)gemrunheader+
and \verb+fill(build)gemeve+ that call the corresponding routines for all the Modules.\\
The \verb+buildgemrunheader+ is called once per run and \verb+buildgemeve+ is called 
once per event to build the banks from the variables in common blocks before calling 
the I/O ZEBRA routines in REM.\\

%% file: database.tex

Run dependent information such as geometry, calibrations and analysis
parameters are stored in a relational database, used for the DAQ
frontend, analysis and simulation. Online data logger inserts
an entry into the database immediately when a run is taken. A run can be
processed by Analyzer with the default settings and reprocessed later with
improved calibration constants after modifying the database. For simulation, the
dedicated program DB2Cards reads the database and write the
FFREAD cards required by GEM for all the configurations required.
Therefore all packages use consistently a common database.

For the main database, MySQL \cite{mysql} is used so that clients can connect over the
network. Daily snapshots are taken in MySQL script format and SQLite\cite{sqlite} format.
SQLite is a single file database; therefore it can be used without network, and can be
used for test purposes by modifying local copies without affecting other users.
Information on all the runs and all the simulation configurations are stored
in the database.
The MEG database consists of a few hundreds tables and each has a direct or indirect
relation to the mother table \verb+RunCatalog+ so that a run number suffices
to retrieve all the information, and no recompilation or manual modification 
of configuration files is required to analyze any run sample. At May 2011, the 
size of the MySQL database is $\approx 500$ MB.

%% file: rome.tex

\begin{figure}[htbp]
  \begin{center}
    \begin{tabular}{cc}
                     \includegraphics[width=.56\linewidth]{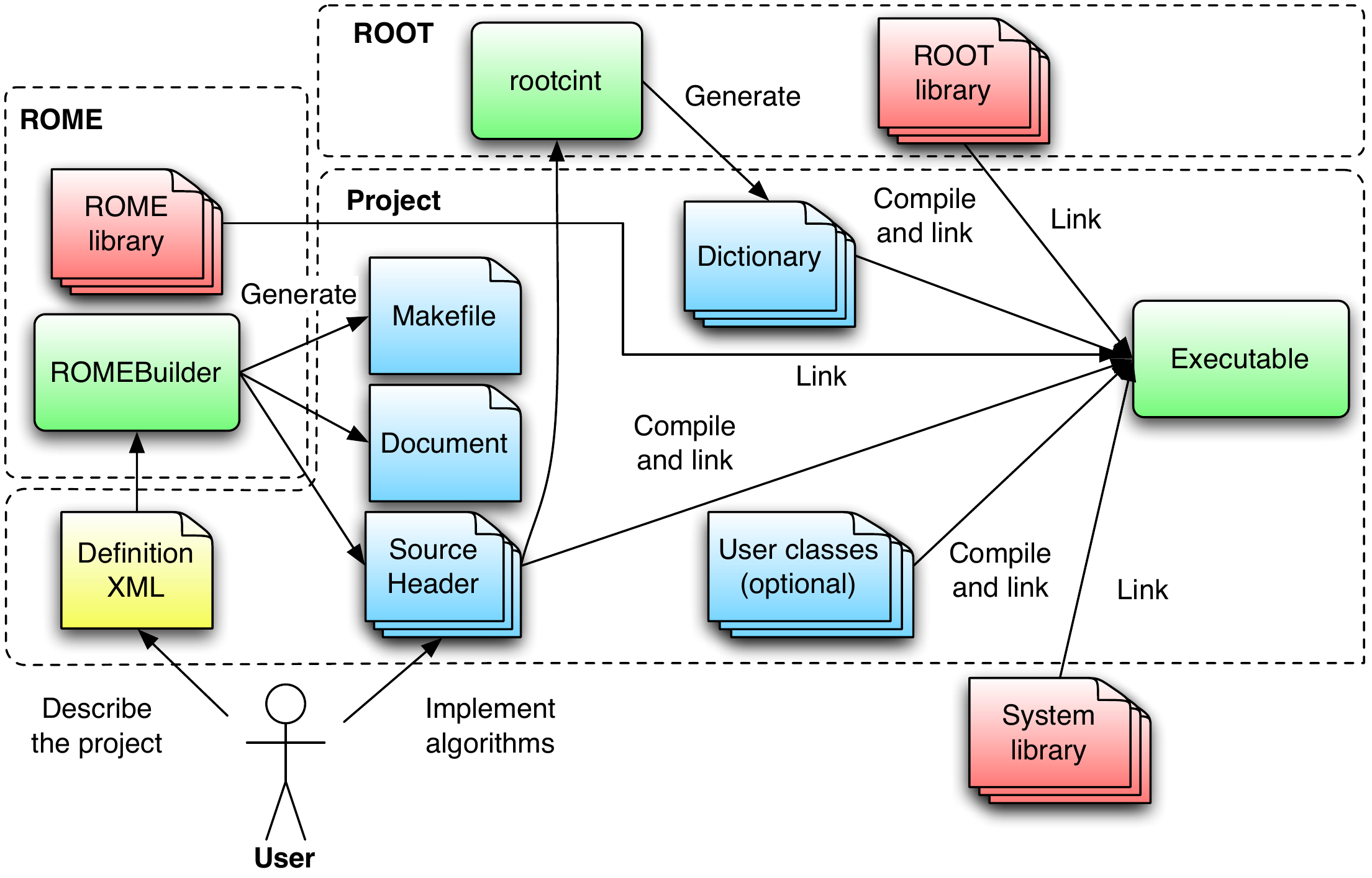} &
    \raisebox{1.0cm}{\includegraphics[width=.39\linewidth]{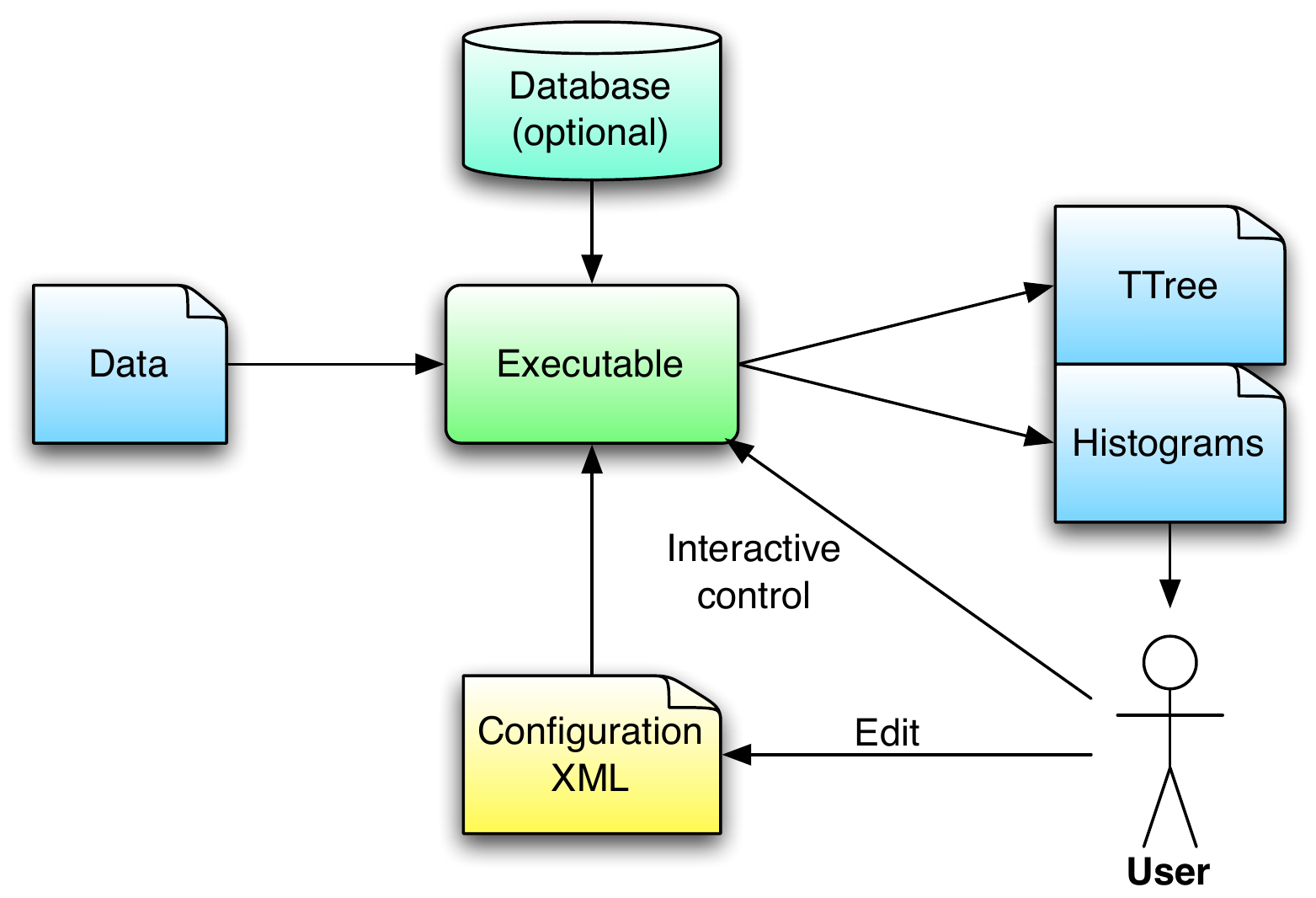}}\\
   (a) & (b)
    \end{tabular}
  \caption{Components in the ROME environment (a) at build time, and (b) at run time.}
  \label{rome_environment}
  \end{center}
\end{figure}

ROME \cite{rome},\cite{romechep06} is a ROOT based framework generator for event based
data processing. 
It has been developed in the MEG collaboration but has been designed as a general-purpose
software so that it can be used for other experiments too.\\
The key concept of ROME is to generate most of the code of a project, except 
the analysis (or simulation) algorithms.\\
In general, data processing software consist of three parts: the first is a project independent 
part such as e.g. user interface, handling of the event loop. The second is a 
project dependent part, which can be summarized in a compact way such as e.g. data
structure and calling sequence of algorithms.
The third is a completely project dependent part such as e.g. the implementation 
of analysis algorithms.\\
Figure \ref{rome_environment} shows components in the ROME environment.
In this environment, the first part is included in the ROME package, and also the ROOT
infrastructure is used. For the second
part, a programmer describes the framework for his/her experiment in a clear and compact 
way in a XML definition file. Out of this file, \verb+ROMEBuilder+ program generates all experiment 
specific classes and modifies the framework. It calls also \verb+make+
command after the source code generation; therefore the build procedure shown in Fig.~\ref{rome_environment}-(a) can be done with a
single command. 
For the third part, a programmer adds the algorithm code to the pre-generated methods.
Further modifications can be done by editing the definition XML file or by modifying algorithm
implementations, then running \verb+ROMEBuilder+ again.\\
Because of the generation scheme, amount of hand written code becomes smaller, and it
becomes possible to start or modify software without learning complicated implementation
of the framework.\\
The generated framework is linked with the ROOT libraries; therefore all ROOT classes 
are available for the analysis. Additional classes written by hand can be also linked.
The generated program is steered using a configuration XML file at the run time.
Interactive control of the program, for example pausing the event loop and ploting histograms, is possible.\\

The following list is part of the items automatically generated by ROME according to a
XML definition file.

\begin{itemize}
\item Data classes (\verb+Folder+s), with a complete set of methods.
\item Algorithm classes (\verb+Task+s) with empty methods to be filled by a programmer.
\item Visualization classes (\verb+Tabs+) with empty methods to be filled by a programmer.
\item Data input classes to read user defined data files with empty methods to be filled by a programmer.
\item Code to create and write histograms. The histograms can be filled in user code.
\item Code for I/O of \verb+TTree+s\footnote{TTree is the ROOT implementation of 
the data structure tree concept} into files.
\item Code to read and write configuration XML files.
\item Code to read and write SQL database. MySQL, PostgreSQL \cite{postgresql} and SQLite are supported and
switchable by a configuration file at run-time.
\item Code to read MIDAS format files and to connect to MIDAS Online Database System (ODB) to access online data.
\item Makefile is automatically generated or updated when new classes are defined by a definition XML.
\item HTML document where description of \verb+Task+s and that of each variable in \verb+Folder+s are
written. ROOT style document, like ``reference guide'' in ROOT web page can be also generated for user code.
\end{itemize}

ROME implements the organization commonly used in OO applications in high energy
physics \cite{lhcb}: {\it data objects}, whose function is to store data, 
are separated from {\it algorithm objects}, whose function is to incorporate algorithms.\\
The former are implemented as a \verb+Folder+ class, the latter as a \verb+Task+ class. 
\verb+Task+s are derived from ROOT \verb+TTask+; therefore recursive calling sequence is realized.
ROME \verb+Folder+s are derived from ROOT \verb+TObject+ (not from \verb+TFolder+), and 
they can be filled into ROOT \verb+TTree+ as a single object or as an array in ROOT \verb+TClonesArray+.\\
For \verb+Folder+s, ROME generates not only the class itself, but also modifies the part of the 
framework related to the \verb+Folder+ such as allocation and initialization, adding or setting 
address of a branch in a \verb+TTree+ for writing (reading) the \verb+Folder+ to (from) a file,
filling variables by reading the database at the beginning of a run (if required in XML).\\
A definition of a \verb+Folder+ reads like a XML document shown in Fig.~\ref{folder_example} 
together with part of C++ code generated by ROME according to the definition.
This \verb+Photon+ instance has two variables, \verb+Energy+ and \verb+Time+.
The generated class has these variables as its data members, and \verb+Set+ and \verb+Get+ 
methods are defined.
The framework generates automatically, for example, 10 instances (the number can 
be fixed or variable) at the beginning of the program and those instances are available in the 
user code out-of-package. For example, \verb+GetPhotonAt()+ and \verb+GetEnergy()+ shown in 
Fig.~\ref{task_example} are generated according to the description in the XML definition file of 
the \verb+Folder+ without manual programming.
Any types of \verb+Field+, both fundamental and derived, can be added in the \verb+Folder+ 
structure as far as it is supported by ROOT dictionary generation (dictionaries are needed
for \verb+TTree+ I/O or socket connection over the network). 

\begin{figure}[htbp]
  \begin{center}
\begin{lstlisting}[language=XML,frame=single,basicstyle=\footnotesize\ttfamily,stringstyle=\rmfamily\itshape,identifierstyle=\rmfamily\itshape]
<Folder>
   <FolderName>Photon</FolderName>
      <ArraySize>10</ArraySize>
      <Field>
         <FieldName>Energy</FieldName>
         <FieldType>Double_t</FieldType>
         <FieldComment>Energy of a photon</FieldComment>
      </Field>
      <Field>
         <FieldName>Time</FieldName>
         <FieldType>Double_t</FieldType>
         <FieldComment>Time of a photon</FieldComment>
      </Field>
</Folder>
\end{lstlisting}

\begin{lstlisting}[language=C++,frame=single,basicstyle=\footnotesize,stringstyle=\ttfamily,identifierstyle=\ttfamily]
class MEGPhoton : public TObject
{
protected:
   Double_t  Energy;  // Energy of a photon
   Double_t  Time;    // Time of a photon
...

public:
   MEGPhoton(Double_t EnergyV=0, Double_t TimeV=0);
   virtual ~MEGPhoton();
...
   Double_t GetEnergy() const         
               { return Energy; }
   Double_t GetTime() const           
               { return Time;   }
   void  SetEnergy(Double_t Energy_v) 
               { Energy = Energy_v; }
   void  SetTime(Double_t Time_v)     
               { Time   = Time_v;   }
...
}
\end{lstlisting}
\caption{An example of \texttt{Folder} definition in a XML file (upper) and part of
   the C++ code generated by ROME (lower).}
\label{folder_example}
  \end{center}
\end{figure}

A definition of an algorithm object, that is a \verb+Task+, reads like a XML document shown 
in Fig.~\ref{task_example}.
According to the definition file, ROME generates header and source files.
A generated source file has empty methods, and a programmer can implement analysis 
in it immediately. As an example, in the code in Fig.~\ref{task_example}, a few lines to access a
\verb+Folder+ are added to the generated file.
ROME generates not only the task class itself, but modifies framework to call it in an
order specified in the definition XML.
In this example, a configuration parameter \verb+DebugPrint+ can be changed using a
configuration XML file at run-time without re-compile.
A function call \verb+GetSP()->GetDebugPrint()+ shown in the example code is available without any
manual programming, and a field to configure the parameter automatically appears in a
configuration XML file after the first use of the file.

\begin{figure}[htbp]
  \begin{center}
\begin{lstlisting}[language=XML,frame=single,basicstyle=\footnotesize\ttfamily,stringstyle=\rmfamily\itshape,identifierstyle=\rmfamily\itshape]
<Task>
   <TaskName>PhotonAnalysis</TaskName>
   <SteeringParameters>
      <SteeringParameterField>
         <SPFieldName>DebugPrint</SPFieldName>
         <SPFieldType>Bool_t</SPFieldType>
      </SteeringParameterField>
   </SteeringParameters>
</Task>
\end{lstlisting}

\begin{lstlisting}[language=C++,frame=single,basicstyle=\footnotesize,stringstyle=\ttfamily,identifierstyle=\ttfamily]
...
void MEGTPhotonAnalysis::Init()
{
}

void MEGTPhotonAnalysis::BeginOfRun()
{
}

void MEGTPhotonAnalysis::Event()
{
...
  if (GetSP()->GetDebugPrint()) {
    for (int i=0;i<10;i++) {
      cout
        <<gAnalyzer->GetPhotonAt(i)->GetEnergy()
        <<endl;
    }
  }
...
}

void MEGTPhotonAnalysis::EndOfRun()
{
}

void MEGTPhotonAnalysis::Terminate()
{
}
...
\end{lstlisting}
\caption{An example of \texttt{Task} definition in a XML file (upper), and part of
   the C++ code generated by ROME (lower).}
\label{task_example}
  \end{center}
\end{figure}

The framework outputs one or more \verb+TTree+s. A programmer can define \verb+Tree+s and 
add \verb+Folder+s to it as branches in a XML description file.
The framework code is automatically modified; therefore no manual programming is needed to add branches to be read or written.
Figure \ref{tree_example} shows an example of \verb+Tree+ structure.\\
Output files can be used for interactive analysis, and further analyzed 
by ROOT macros.

\begin{figure}[htbp]
  \begin{center}
\begin{lstlisting}[language=XML,frame=single,basicstyle=\footnotesize\ttfamily,stringstyle=\rmfamily\itshape,identifierstyle=\rmfamily\itshape]
<Tree>
   <TreeName>DataTree</TreeName>
   <Branch>
      <BranchName>PhotonBranch</BranchName>
      <RelatedFolder>Photon</RelatedFolder>
   </Branch>
</Tree>
\end{lstlisting}
\caption{An example of \texttt{Tree} definition in a XML file.}
\label{tree_example}
\end{center}
\end{figure}

Output files of each step can be used as input files of the following step; therefore the
analysis can be separated into several steps.
For example, in the analysis of MEG, we can save results of waveform analysis, which is
the most time-consuming in the chain, and perform reconstructions on this file to improve
the algorithm many times without redoing the waveform analysis.

An interactive mode, which is almost the same as ROOT interactive mode, is also provided.
In the interactive mode or in macros, experiment specific classes are
also available in addition to the standard ROOT classes.\\
ROME also generates a HTML document and a Makefile. The generated framework is
already compilable just by \verb+make+ command and, after that, is executable.\\
The generation mechanism is used not only at the beginning of the project, but also during
the code development. For example, a programmer can easily add a new configuration parameter to an
existing \verb+Task+, or add new variables to a \verb+Folder+. Code in the framework is automatically
modified consistently.\\
MEG Analyzer consists of about 200 \verb+Folder+ classes and 100 \verb+Task+ classes. 
The total number of lines in the Analyzer code is more than one million.
84\% of them are either generated by \verb+rootcint+ \cite{root} or ROME, or included 
in the ROME package, while the rest were written manually.

%% file: bartender.tex
Following the detector simulation and before the reconstruction and analysis program an 
intermediate program, called Bartender, is required for the processing of Monte Carlo data.
This program serves different roles:

\begin{itemize}
\item Conversion of ZEBRA files into ROOT files
\item Readout simulation
\item Event mixing
\end{itemize}

It reads the GEM output ZEBRA files calling \verb+fillgemrunheader+ once per run and
\verb+fillgemeve+ once per event after calling the I/O ZEBRA routines to fill the
variables in FORTRAN common blocks from the banks. These variables are finally
mapped to C++ classes manually.\\
Simulation specific data such as kinematics of generated particles, 
true hit information, etc.~can be streamed in a \verb+sim+ \verb+Tree+ in separate ROOT 
files for further studies.

It simulates detector readout electronics and produces waveforms. For
example, the Liquid Xenon Calorimeter waveforms are obtained by convolution of single
photoelectron response of a photomultiplier tube (PMT) with hit-time information of scintillation
photons simulated in GEM. PMT amplification, signal attenuation, saturation of the
readout electronics, noise, etc.~are taken into account. Simulated waveforms are
encoded in the same manner as the experimental ones and written in a \verb+raw+ 
\verb+Tree+ in ROOT files. 

It makes a mixture of several sub-events; rates of each event type 
are set with a configuration file.
To study the combinatorial background events,
sub-events are mixed with various relative timing with respect to each other
and with respect to the trigger. For instance random and fixed timing can be selected.
That allows simulating many different pile-up configurations with a limited number 
of samples of events simulated through the detector.

%% file: analyzer.tex
Analyzer incorporates multiple purposes: event reconstruction,
visualization, computation of calibration constants and physics analysis.

\subsection{Event reconstruction}
Analyzer consists of several \verb+Task+s for each step of analysis;
each \verb+Task+ can be switched on/off.\\
In the first step, raw data are read and calibrations are applied to waveforms.
In the second step, waveform analysis specialized for each sub-detector are 
performed to extract time and charge of pulses. Waveforms are also used to 
identify pileup events and for particle identifications.\\
In the third and last step, events are reconstructed using algorithms implemented by
experts of each sub-detector. Several different algorithms are implemented 
to reconstruct each kinematic parameter for crosschecks. Each \verb+Task+ may have
a dedicated \verb+Folder+ to write its result. \verb+Task+s share a \verb+Folder+ to hold results of
a standard choice among those algorithms; this choice is specified by a configuration file.
\verb+Task+s are executed in the same process and results are written in an 
output file together.

\begin{figure}[htbp]
  \begin{center}
    \includegraphics[width=.95\linewidth]{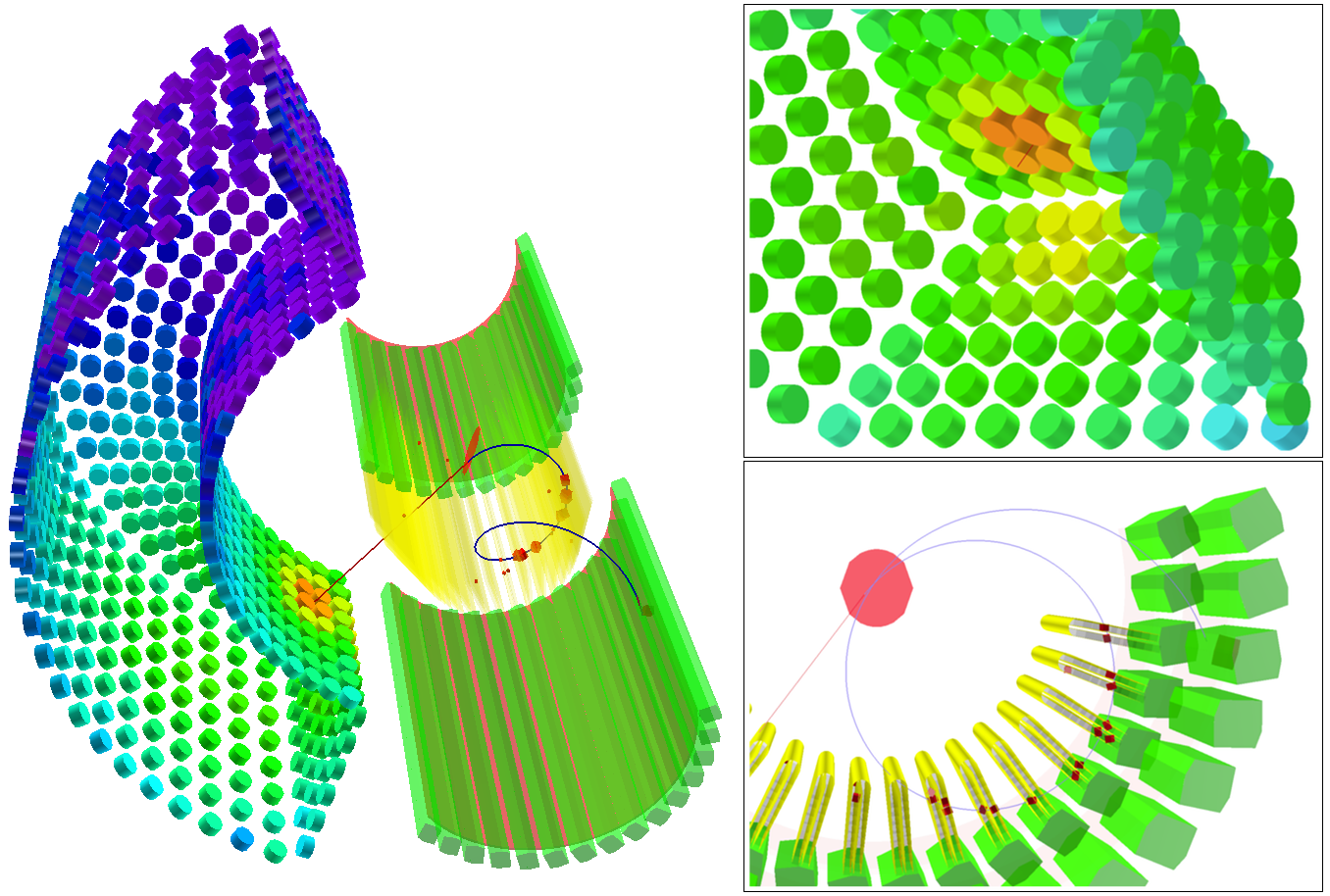}
    \caption{A $\mu^+\to e^+\gamma$ reconstructed event and closer views. 
     Reconstructed hits in drift chambers and timing
       counters, a positron track and a $\gamma$-ray are shown. Color-code of Calorimeter
    PMTs represents output of each PMT.}
    \label{eventdisplay}
  \end{center}
\end{figure}

Figure \ref{eventdisplay} shows a reconstructed experimental event.

\subsection{Visualization}
Data quality is monitored for various time-spans: event-by-event, run-by-run or in days.

\begin{figure}[htbp]
  \begin{center}
    \includegraphics[width=.95\linewidth]{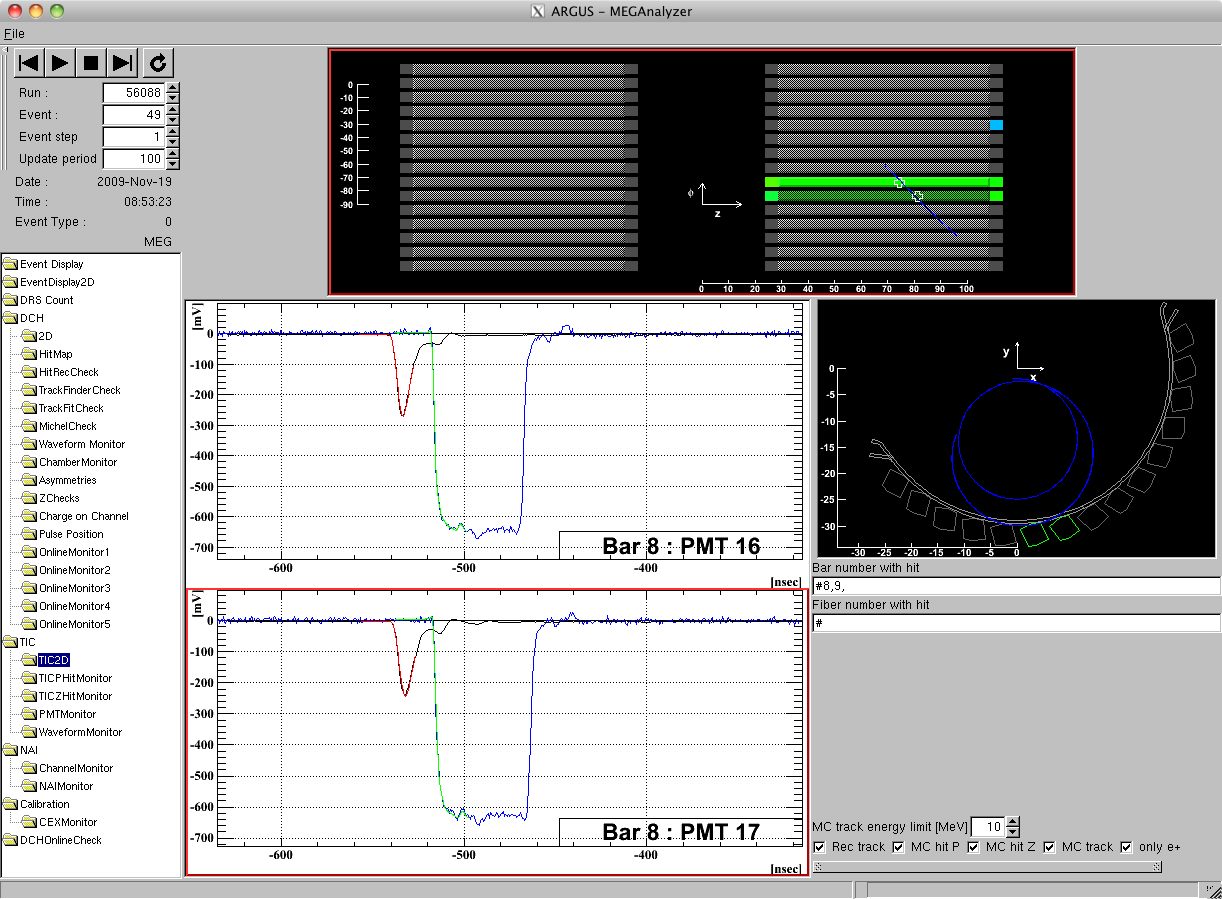}
    \caption{A graphical display of timing counter hits, waveforms. A reconstructed positron track is also shown.}
    \label{display}
  \end{center}
\end{figure}

For event-by-event monitoring, several displays are implemented. Figure \ref{display}
shows one of them. The displays show waveforms, status of trigger, reconstructed hits and tracks
and any other information useful for monitoring. Those displays are used for both online and offline.
When it is used for online monitoring, Analyzer and DAQ run in parallel and data are
transferred over a socket connection. Hard copies of the displays are saved periodically
for remotely monitoring using web-browsers.

Two types of portable document format (PDF) files are automatically prepared by macros,
    which read histogram files made by Analyzer.
The first type shows histograms to describe
the run and is made automatically for each run soon after the run is finished.
The second type shows strip charts to monitor time variations of the status 
of the detector and of the electronics in a day or a week.

\subsection{Calibration}
Analyzer is used also to compute calibration constants (photomultiplier gains, time-offsets, etc.).
Each calibration constant is associated to a \verb+Task+. The calibration \verb+Task+s are usually run on 
events already processed with a preliminary set of calibration constants. The updated 
calibration constants can be made available in a variety of format: histograms, text file
or SQL macro.
They can be stored in the database, and used in the next round of reconstruction.

\subsection{Physics analysis}
Event preselection and blinding for physics analysis, described in
section \ref{sec_offline}, are implemented in Analyzer.
On events in the analysis region, likelihood analysis is performed to
calculate the best estimate of the number of $\mu^+\rightarrow e^+\gamma$ signal candidates, its
confidence interval and the significance.
The 90\% confidence interval of the number of signal events is calculated using the unified
approach \cite{feldman_1998}.
We made independent likelihood analysis tools with different statistical methods
or parametrization of probability density functions for cross checks.

%% file: offline.tex
Just after a run is taken, a raw data file written in the MIDAS format is sent to the
offline-cluster, and a process to analyze it automatically starts.
The MIDAS files are compressed and stored on tapes.
The compressed MIDAS files and \verb+rec+ files of calibration runs are accessible for further
studies, while a special treatment is done for the data of physics runs.\\
MEG has adopted the principle of 'blind' analysis in searching 
for $\mu^+\rightarrow e^+\gamma$ signal. That means that the events 
with kinematic parameters closest to the expected signal (in the 'blinding box')
cannot be used for determining the analysis parameters (e.g. the 
cuts or the probability density functions) to avoid biasing the analysis. 
In order to guarantee that, the data in the 'blinding box' are 
inaccessible during the first phase of the analysis.\\
This concept is realized in Analyzer with \verb+Task+s 
streaming the events into different ROOT files depending on the 
selection criteria they satisfy.\\
A first round of processing operates a pre-selection on coarsely 
calibrated data with loose cuts that are streamed in:
\begin{basedescript}{\desclabelstyle{\pushlabel}\desclabelwidth{6em}}
\item[selected]  Events passing the pre-selection
\item[unselected] Events not passing the pre-selection
\item[unbiased] All calibration trigger events and every fiftieth physics-trigger event  
\end{basedescript}
\verb+Trees+ containing raw waveforms are produced for 'selected' and 'unbiased' events in this step.
The 'unbiased' samples are used for monitoring of the experiment and for the calibrations.
The 'selected' events are not accessible.
After the calibrations are finalized, reconstruction is performed on the 'selected' samples
using \verb+raw+ files. At the end of this step, another \verb+Task+ applies tighter cuts defining the 'blinding box'.
The events are streamed into the files:
\begin{basedescript}{\desclabelstyle{\pushlabel}\desclabelwidth{6em}}
\item[blind] Events preselected in the 'blinding box', candidate to be signal
\item[open]  Events preselected but outside the 'blinding box'
\end{basedescript}
and 'selected' files are deleted.
The 'blind' files are made accessible only when the analysis is finalized.

%% file: conclusion.tex
Software is a crucial component of any experiment and its power and flexibility is a key
ingredient of its success.\\
MEG had the challenge to design a software structure that could strike a balance
between flexibility and user friendliness. The limited size of the offline 
architecture group and the requirement that a large fraction of the
collaboration could contribute to the programming of the algorithms, 
have led to greatly emphasize the use of known packages as well as the
shielding from the average programmer of I/O handling, format conversion and 
Object Oriented programming into the frameworks.\\
A mixed language environment with two separate frameworks, one for each environment,
proved to be successful. 
It relies heavily on standard software elements like GEANT3, ZEBRA, 
FFREAD in the simulation section implemented in FORTRAN 77; XML, ROOT in the rest of the code 
implemented in C++; MySQL and SQLite for the database.\\
This configuration allowed the implementation of all experimental requirements within the tight
time and manpower constraints, such to support the physics analysis first published in 
\cite{meg2009}.